\documentclass{rspublic}
\usepackage[dvips]{graphicx}
\usepackage{natbib}
     \bibpunct{(}{)}{;}{a}{}{,}

\newcommand{\BE}{\begin{equation}}
\newcommand{\EE}{\end{equation}}
\newcommand{\BA}{\begin{eqnarray}}
\newcommand{\EA}{\end{eqnarray}}

\begin{document}

\title[Species competition]{Species competition:
coexistence, exclusion and clustering}

\author[E. Hern{\'a}ndez-Garc{\'\i}a, C. L{\'o}pez, S. Pigolotti and K.H. Andersen]
{Emilio Hern\'andez-Garc{\'\i}a$^1$, Crist{\'o}bal L{\'o}pez$^1$,
Simone Pigolotti$^2$, Ken H. Andersen$^3$}

\affiliation{$^1$IFISC (UIB-CSIC), Instituto de F{\'\i}sica
Interdisciplinar y Sistemas Complejos, Campus Universitat de les
Illes Balears, E-07122 Palma de Mallorca, Spain
\\$^2$The Niels Bohr Institute, The Niels Bohr International
Academy,  Blegdamsvej 17 DK-2100, Copenhagen, Denmark
\\$^3$National Institute of Aquatic Resources, Technical University of
Denmark, Charlottenlund slot, DK-2920 Charlottenlund, Denmark}

\label{firstpage}
\maketitle

\begin{abstract}{Competition, Lotka-Volterra,
competitive exclusion, limiting similarity, pattern formation} We
present properties of Lotka-Volterra equations describing
ecological competition among a large number of interacting
species. First we extend previous stability conditions to the case
of a non-homogeneous niche space, i.e.~that of a carrying capacity
depending on the species trait. Second, we discuss mechanisms
leading to species clustering and obtain an analytical solution
for a state with a lumped species distribution for a specific
instance of the system. We also discuss how realistic ecological
interactions may result in different types of competition
coefficients.\\
\ \\
Philosophical Transactions of the Royal Society A 367, 3183-3195
(2009)\\
http://dx.doi.org/10.1098/rsta.2009.0086\\
Published under the Creative Commons Attribution license. 
\end{abstract}

\section{Lotka-Volterra competition and species distribution}

Competitive interactions occur when entities in a system grow by
consuming common finite resources. They are ubiquitous in many fields
of science: examples include biological species competing for food
\citep{MacArthur1967,Roughgarden1979,Case1981}, mode competition in
nonlinear optical systems \citep{Benkert1991}, or alternative
technologies competing for a market \citep{Pistorius1997}.  An early,
simple, but powerful model for competitive interactions is the
Lotka-Volterra (LV) set of competition equations
\citep{Volterra1926,Lotka1932}:

\begin{equation}
  \label{LV}
  \dot{N_i} = r_i N_i \left(1 - \frac{1}{K_i} \sum_{j =1}^m G_{ij} N_j\right),
  \ \ i=1,...,m.
\end{equation}
where $m$ is the number of species, $N_i$ the population of species
$i$, $r_i$ its maximum growth rate, $K_i$ its carrying capacity, and
$G_{ij}$ is the matrix characterizing the interaction among species
$i$ and $j$, more specifically the decreasing on the growth rate of
species $i$ by the presence of $j$. Competitive interactions are
characterized by $G_{ij} \ge 0$, the situation to be considered here,
whereas negative interactions may model situations of mutualism,
predation or symbiosis.

In classical ecological niche theory, species are associated to
points in an abstract {\sl niche space}. Coordinates in this space
represent relevant phenotypic characteristics, for example size of
individuals in a species, or the size of preferred prey, such that
intensity of competition is larger if species are closer in this
space. We assume for simplicity this space to be one-dimensional
(multi-dimensional generalizations are straightforward, as briefly
mentioned later). If niche locations can be considered to be a
continuum, we can write Eq. (\ref{LV}) as:
\BE
\partial_t \psi (u,t) =   r(u) \psi(u,t)  \left[1
- \frac{1}{K(u)}\int G(u,v)\psi(v,t)dv\right] ,\ \ \ \ \
\label{LVcon}
\EE
where now $\psi(u,t)$ is the population density at niche location
$u$. The integral extends over the full niche space, which could
be finite or infinity. For most purposes, Eqs. (\ref{LV}) and
(\ref{LVcon}) can be considered as equivalent, since the second is
obtained from the first in the limit of many close interacting
species, and (\ref{LV}) can be recovered from (\ref{LVcon}) for a
discrete distribution of species:
\BE
\psi(u)=\sum_{i=1}^m N_i \delta(u-u_i),
\label{deltas}
\EE
with $G_{ij}=G(u_i,u_j)$, $r_i=r(u_i)$ and $K_i=K(u_i)$.

It is widely believed that (\ref{LV}) or (\ref{LVcon}) predict a
{\sl competitive exclusion} leading to a {\sl limiting similarity}
situation \citep{Abrams1983}, in which a pair of species too close
in niche space can not coexist, and one of them would become
extinct. However it is known that the model allows for continuous
coexistence of species in some situations \citep{Roughgarden1979},
and refinements on the conditions for this coexistence have been
developed, with emphasis on the effect of the shape of the
carrying capacity function $K(u)$ \citep{Meszena2006,Szabo2006}.
In this context, a particulary surprising result was the finding
by \citet{Scheffer2006} of a situation --for uniform carrying
capacity-- which was neither of full coexistence nor of full
exclusion, but of clusters or lumps of tightly packed species
which did not exclude each other, but were well separated from
other clusters so that there was a type of limiting similarity
leading to a minimum intercluster distance. Clustering of
individuals or entities under competitive interactions of the LV
type had been already observed in other contexts
\citep{Fuentes2003,HernandezGarcia2004,HernandezGarcia2005jphysc,Ramos2008},
where the mechanism was the diffusive broadening of an otherwise
zero-width species or entity. In contrast, the lumps in
\citet{Scheffer2006} appeared even in the absence of diffusion in
niche space, which is the situation also considered here.

The importance of the functional form of the interaction kernel
$G_{ij}$ in (\ref{LV}) or $G(u,v)$ in (\ref{LVcon}) was stressed
by \citet{Pigolotti2007} for the case of uniform carrying capacity
and interactions depending only on differences of niche positions,
and found to be relevant in an evolutionary context by
\citet{Leimar2008}. For that case the positive-definite character
of the Fourier transform of $G(u,v)=G(u-v)$ is a condition
implying the absence of limiting similarity. Species clustering
was reported, but for interaction functions rather different from
the Gaussian used in \citet{Scheffer2006}. In fact, for the
Gaussian interaction case most results are extremely sensitive to
details such as the implementation of the boundary conditions or
weak ecological second order effects \citep{Pigolotti2008}. Thus,
a clarification of the mechanisms leading to species clustering in
LV models would be desirable. In addition, the results in
\citet{Pigolotti2007,Pigolotti2008} were obtained under the
unrealistic assumption of homogeneity in niche space whereas the
inhomogeneities in the carrying capacity are known to play
relevant roles \citep{Szabo2006}. For simplicity we restrict our
description to the standard situation in which competition is
stronger among species closer in niche space. It is worth
mentioning the existence of studies of LV systems where non-local
interactions are considered \citep{Doebeli2000}. That situation
can also be described by the general formalism used here of an
integral kernel function, and our general results therefore also
apply to the situation with non-local interactions.

In this Paper we analyse some mathematical properties of the LV
model (\ref{LV}) or (\ref{LVcon}). In Sect. \ref{sec:stability} we
show that the positive-definiteness of the kernel $G$ remains a
determining condition for stable coexistence even for non constant
$K(u)$. In Sect. \ref{sec:lumps}, we discuss the mechanism
producing lumped species distributions and explicitly give an
analytic expression for a particular interaction kernel. In the
Appendix we show that, in contrast with the earliest
characterizations of the interaction kernel $G$
\citep{MacArthur1967,Roughgarden1979}, both positive- and
non-positive-definite kernels can arise from more detailed
ecological models which consider the dynamics of the consumed
resource. We use periodic boundary conditions in our numerical
simulations. We expect the effects of this simplifying but
unrealistic assumption to be unimportant at least when a
non-constant carrying capacity limits the presence of species to a
limited region of niche space.

\section{The stability of close coexistence}
\label{sec:stability}

A simplifying assumption for the study of the LV model is that of
homogeneity in niche space. In this case, the carrying capacity
and growth rate are constants, $K_0$ and $r_0$, and the
interaction kernel depends only on differences of niche positions
$G(u,v)=G(|u-v|)$. Niche space could be infinite, but in the case
in which it is finite, homogeneity can only be achieved under
periodic boundary conditions. Under these restrictions it is easy
to see that a steady solution to (\ref{LVcon}) which is
homogeneous and everywhere non-vanishing always exists:
$\psi_0=K_0/\hat G_0$, where $\hat G_0 \equiv \int du G(u)$. This
solution represents coexistence of all possible species without a
limit to their similarity. Its stability against small
perturbations can be analysed by linearization of the equation
resulting from substitution of $\psi(u,t)=\psi_0+\delta\psi(u,t)$
into (\ref{LVcon}). The solution for the Fourier transform of the
deviation from the homogeneous state, $\delta\hat\psi_q(t)$, is
\BE \delta\hat\psi_q(t)=\delta\hat\psi_q(0)
e^{\lambda_q t}\ , {\rm with }\ \ \lambda_q = -r_0\frac{\hat G_q}{\hat
  G_0} \ .  \EE
where $\hat G_q$ is the Fourier transform of $G(u)$. Thus, the
homogeneous solution $\psi_0$ is stable if $\hat G_q$ is positive
$\forall q$, while a instability leading to pattern formation
occurs when $\hat G_q$ may take negative values
\citep{Pigolotti2007,Fuentes2004,HernandezGarcia2004,Lopez2004}.
We note that many steady solutions to Eq. (\ref{LVcon}) exist
besides $\psi_0$ (in particular, solutions of the form
(\ref{deltas})). This is so because dynamics preserves $\psi(u)=0$
at all places where there is no initial population. Notice also
that $\psi_0$ is the only strictly positive solution.  Among this
multiplicity of solutions the ones that will be more relevant are
the ones which are stable under perturbations or small immigration
\citep{Pigolotti2007}.

An interesting class of functions to be used as kernels and
carrying capacities is the family $\{g_\sigma^p\}$ given by
\BE
g_\sigma^p(u) \equiv  \exp\left(-|u/\sigma|^p\right),
\label{Gp}
\EE
which is parameterized by the value of $p$. The widely used
Gaussian kernel is obtained for $p=2$. When $p < 2$ the functions
are more peaked around $u=0$ (the case $p=1$ is an exponential)
and for $p>2$ they become more box-like ($g_\sigma^\infty(u)$ is
the flat box with value $1$ in the interval $[-\sigma,\sigma]$ and
zero outside). The width of the kernel $\sigma$ gives the
competition range in niche space. We have positivity of the
Fourier transform if $p\le 2$. This implies that the homogeneous
solution is stable under evolution with uniform $K$ and kernel $G$
of the form (\ref{Gp}) if $p\le 2$. When $p>2$, the homogeneous
solution is unstable and the system approaches delta comb
solutions of the type (\ref{deltas}), with a spacing approximately
$1.4\sigma$ \citep{Pigolotti2007} which represent limiting
similarity situations.

We now generalize the above stability analysis to the more
realistic case in which there is no homogeneity in niche space.
First we consider the simpler case of a symmetric kernel
$G(u,v)=G(v,u)$, which in particular includes the previous case of
kernels depending only on species distance: $G(u,v)=G(|u-v|)$.
Note that in this symmetric case one can write Eq. (\ref{LVcon})
in potential form:
\begin{equation}
\partial_t \psi (u,t)=-r(u)\frac{\psi(u,t)}{K(u)}
    \frac{\delta V[\psi]}{\delta \psi (u)},
\label{eqpotencial}
\end{equation}
with the functional potential given by:
\BE
V[\psi] = - \int K(u)\psi(u,t)du +  \frac{1}{2}\int\int
G(u,v)\psi(u,t)\psi(v,t)\ du\ dv.
\label{potencial}
\EE

Stationary solutions of Eq. (\ref{LVcon}) are those for which the
r.h.s of Eq. (\ref{eqpotencial}) equals $0$. This has many
possible solutions. We define the {\em natural stationary solution},
$\psi^N (u)$, as the one which is positive and non vanishing for
all $u$, so that
\begin{equation}
\left(\frac{\delta V}{\delta \psi}\right)_{\psi^N}=0,
\end{equation}
that is, the one satisfying:
\begin{equation}
\int G(u,v)\psi^N (v)dv = K(u) \ , \forall \ u
\label{natural}
\end{equation}
The solution $\psi^N (u)$ can be considered the non-homogeneous
generalization of $\psi_0$ introduced in the homogeneous case.  In the
particular case in which $G(u,v)=G(u-v)$ the natural solution can be
explicitly written in terms of Fourier transforms of the competition
kernel and the carrying capacity, either in an infinite system or in a
finite one with periodic boundary conditions:
\BE
\hat \psi^N_q=\frac{\hat K_q}{\hat G_q}.
\label{eq:FourierNatural}
\EE
This requires that these Fourier transforms and their inverses
exist and lead to positive populations densities.  When this
happens, a continuum species coexistence is obtained, and its
existence is generally robust against small changes in $G$ or $K$.
We show later that it is also an attractor of the dynamics when
$\hat G_q$ satisfy positivity requirements ($p\le 2$, for the
family in (\ref{Gp}), being $p=2$ the marginal case). For a
uniform carrying capacity, the natural solution (9) always exists
and is uniform in phenotype space $\psi^N(u)=\psi_0$. But the
natural solution may lose positivity or even cease to exist
depending on the properties of $G$ and $K$. For example, when both
$G(u)$ and $K(u)$ are of the form (\ref{Gp}) with $p=2$, the
inverse Fourier transform of (\ref{eq:FourierNatural}) exists when
the carrying capacity has a value of $\sigma$ larger than the
kernel $G$, but not in the opposite case.

\begin{figure}[t]
  \centering
 \includegraphics[width=\columnwidth,clip]{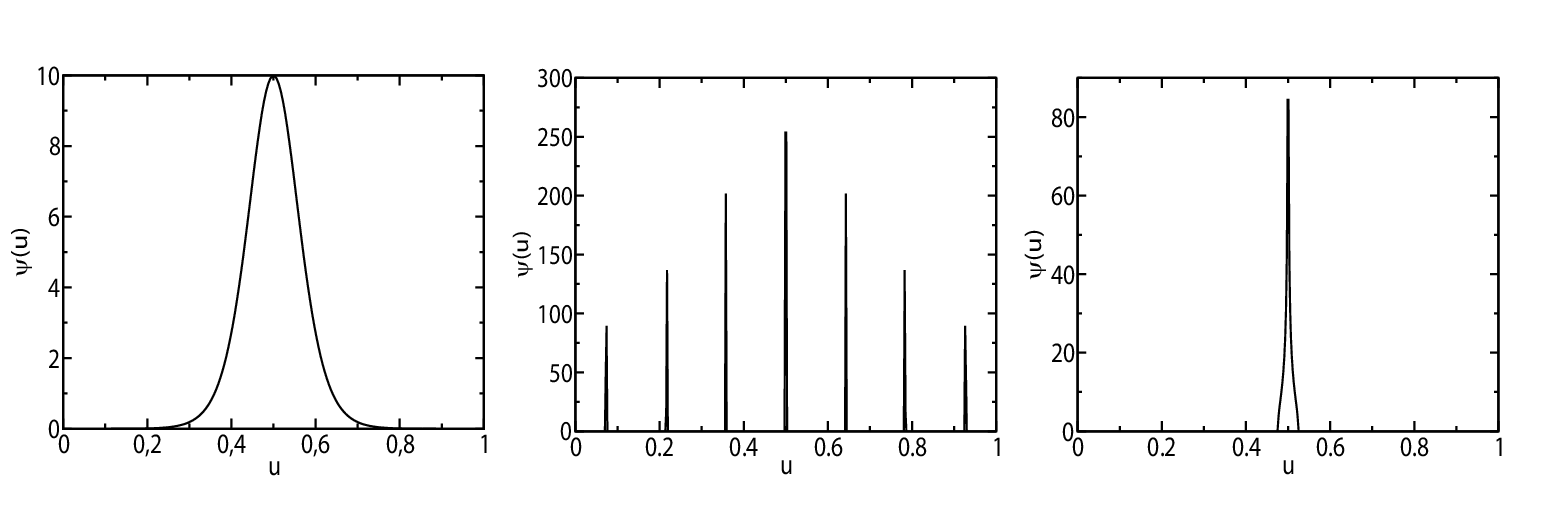}
  \caption{Long-time solutions of (\ref{LVcon}) for different kernels
  and carrying capacities. Left: $G=g_\sigma^1$,
  $K={\rm sech}(u/\sigma)$, with $\sigma=0.1$. The natural steady solution
  ($\psi^N=a^{-1}{\rm sech}^3(u/\sigma)$), which is positive and non-vanishing
  everywhere, is reached at long times. Center: $G=g_{0.1}^4$,
  $K=g_{0.1}^{0.5}$. Under this non-positive-definite competition kernel, the
  solution shown is still evolving and approaches a singular delta comb of the
  type (\ref{deltas}) at long times. Right: $G=g_{0.1}^{0.5}$,
  $K=g_{0.1}^{1}$. A positive natural solution does not exist and the system
  approaches a single hump solution which vanishes in part of niche space.
  }
\label{fig:numerical}
\end{figure}

Figure \ref{fig:numerical} shows stationary solutions attained at
long times by the dynamics in (\ref{LVcon}) illustrating the
situations described above, starting from a smooth non-vanishing
initial condition. In the first case we choose a kernel and
carrying capacity functions ($G(u)=g_\sigma^1(u)$, $K(u)={\rm
sech}(u/\sigma)$) such that the natural solution exists and is
positive everywhere. Thus it is stable, and it is the steady state
attained at long times. In fact it can be analytically calculated:
\BE
\psi^N(u)=a^{-1}{\rm sech}^3(u/\sigma) \ .
\EE

In the second case the non-positiveness of the kernel used (with a
carrying capacity of the type Eq.(\ref{Gp})) breaks down the
initial configuration into lumps, which at long times approach
zero-width delta functions with forbidden zones in between. In the
third case, despite $\hat G_q$ being positive, a positive natural
solution does not exist. Several outcomes are possible but for the
kernel and capacity used, the system approaches a single hump
solution which vanishes in part of the niche space.

More in general, but still in the symmetric $G$ case
$G(u,v)=G(v,u)$, writing the LV model in potential form (Eq.
(\ref{eqpotencial})) is of great use since one can show that,
provided $r(u)$ and $K(u)$ are positive, $dV/dt \leq 0$. This
implies that $V$ is a Lyapunov potential and dynamics proceeds
towards its absolute minimum, or if $\psi(u,t=0)=0$ for some $u$,
towards the minimum of $V$ under such constrain. Notice that,
since the potential is a quadratic form, $\psi^N$ is a {\it
global} attractor (starting from non-vanishing initial conditions)
when the competition kernel is a positive definite quadratic form,
which means that $\int\int f(u)G(u,v)f(v)du dv \ge 0$, $\forall f$
(or $\sum_{ij} x_i G_{ij} x_j \ge 0$, $\forall \{x_i\}$ in the
discrete case). This generalizes the previous stability condition
on the Fourier transform $\hat G_q>0$ to niche inhomogeneous
cases, and shows that the stability result was global indeed. In a
multi-dimensional niche space the same analysis shows that the
positive-definiteness of the quadratic form remains the condition
for the global stability of the natural solution. In any case, the
important consequence is that the stability of the natural
solution depends uniquely on the competition kernel and not on the
carrying capacity (provided the relation kernel-capacity is such
that the natural solution exists and is positive). In particular,
for competition kernels of the form (\ref{Gp}), $\psi^N$ is always
(if existing and positive) a globally stable solution of the
dynamics for $p\le 2$, and unstable otherwise.

The crucial difference in the case of a non-symmetric competition
kernel is that there is no obvious Lyapunov potential for the
system. This implies that there are no
available global stability results. However, local stability can
be investigated. Let us consider a small perturbation of the
positive natural solution $\psi^N (u)+ \delta \psi (u,t)$.  To
linear order, the perturbation evolves as:
\begin{equation}
\frac{d \delta \psi (u,t)}{dt}= -r(u)\frac{\psi^N(u)}{K(u)}\int
G(u,v)\ \delta \psi (v,t)\, dv .
\label{evolpert}
\end{equation}
We now consider the functional $H(\delta \psi) \equiv \int \ du \
\left(A(u)K(u)/r(u)\psi^N(u)\right)\ (\delta \psi)^2$, where
$A(u)$ is a positive function so that $H \ge 0$ and $H(0)=0$. Let
us compute its time derivative:
\begin{equation}
\frac{d H}{dt}=-2  \int  \delta \psi(u) \ A(u)\ G(u,v) \ \delta
\psi(v) \, du\, dv.
\label{Fevol}
\end{equation}

If for some choice of $A(u)$ one has that $A(u) G(u,v)$ is
positive definite, then $dH/dt<0 $ and $\delta \psi =0$ will be
approached. This shows that $\psi^N$ is linearly stable in such
case. We note that the case in which $G(u,v)$ itself is
positive-definite trivially guaranties the positivity of $A(u)
G(u,v)$, with a constant $A$.  Thus, even in this more general
nonsymmetric case, it is the character of the interaction kernel
$G$, and not of the carrying capacity (provided it is such that
the natural solution exists and is positive), which
determines the stability of the natural solution.

\section{Lumped species distributions}
\label{sec:lumps}

\begin{figure}[ht]
    \begin{center}
    \includegraphics[width=\columnwidth,clip]{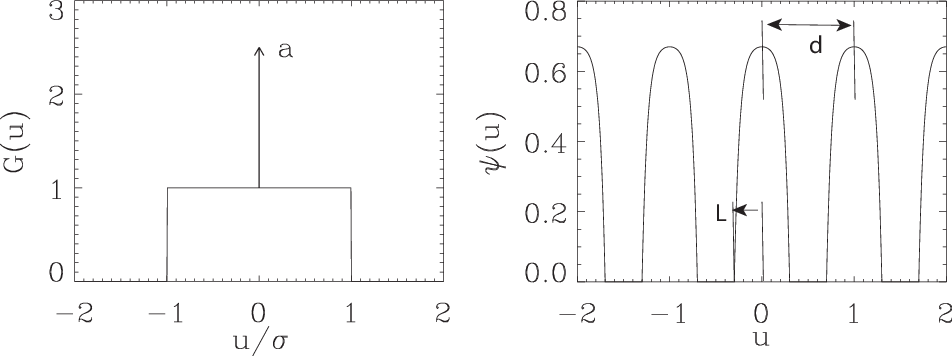}
    \end{center}
\caption{\label{fig:analyticlumps} The kernel in Eq. (\ref{flat+delta})
(left), and the analytic steady solution given by
(\ref{sumoflumps}) and (\ref{fhump}-\ref{normalization}) for
$a=K_0=1$, $\sigma=0.8$, $L=0.3$ and $d=1$ (right).}
\end{figure}

\citet{Scheffer2006} found transient but long-lived solutions of
Eq. (\ref{LV}) consisting of periodically spaced lumps containing
many close species. They used a Gaussian interaction kernel which
turned out to introduce an excessive sensitivity of the results to
the numerical implementation of the model and boundary conditions
\citep{Pigolotti2008}. They found however similar solutions as
steady configurations when adding an extra predation term acting
effectively only on species with high population. This can be
thought as an extra {\sl intraspecific} competition since it
decreases the growth of species with many individuals. Exploiting
this idea, \citet{Pigolotti2007} checked the effect of using in
(\ref{LVcon}) a kernel of the type (\ref{Gp}) but with an enhanced
interaction at $u=0$, i.e. enhanced intraspecific competition. In
particular, they used a constant carrying capacity $K(u)=K_0$ and
a flat box kernel with an added delta function at the origin (see
Fig. \ref{fig:analyticlumps}),
\BE
G(u)=g_\sigma^\infty(u)+a \delta(u) \ .
\label{flat+delta}
\EE
Lumped patterns were obtained numerically for $a=1$.

Because the dynamics of (\ref{LVcon}) usually involves very long
transients, it is interesting to calculate analytically the steady lumped
solution in the simple case of a kernel (\ref{flat+delta}) and uniform
carrying capacity $K_0$ (in the infinite line).

\begin{figure}[ht]
    \begin{center}
    \includegraphics[width=.8\columnwidth,clip]{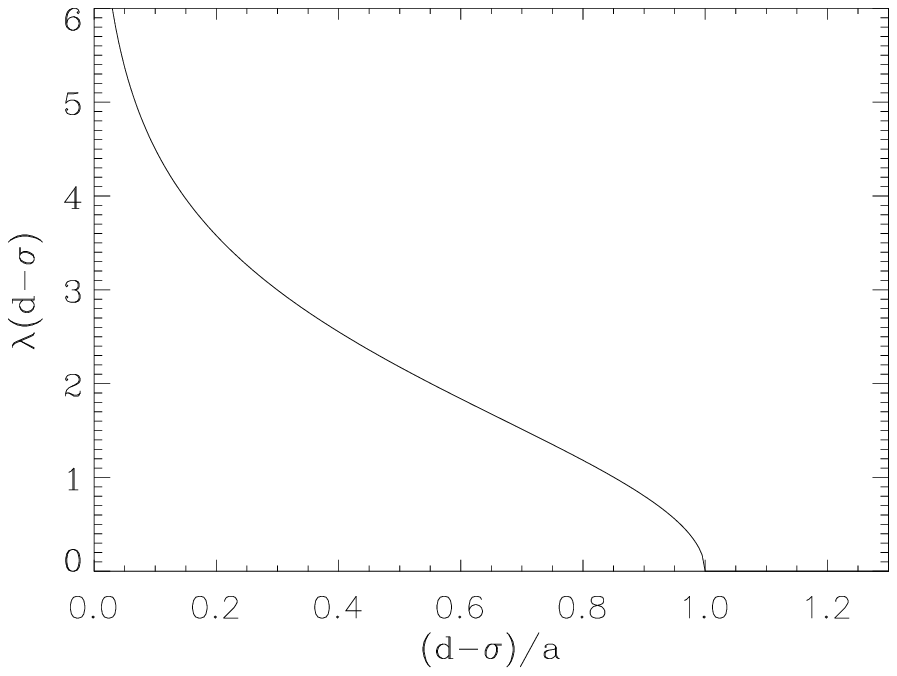}
    \end{center}
\caption{\label{fig:lambda} The solution $\lambda$ (positive branch)
of Eq. (\ref{trascen}), giving the inverse width of species lumps.
The width is finite for $d-\sigma<a$, which is favored by larger
enhanced intraspecific competition $a$. }
\end{figure}

We begin with the steady state condition
\begin{equation}
\int G(u,v)\psi(v)dv = K(u) \ ,
\label{steady}
\end{equation}
valid at $u$ such that $\psi(u)\neq 0$, that particularized to
(\ref{flat+delta}) and constant $K$ reads:
\BE
a\psi(u)+\int_{u-\sigma}^{u+\sigma} dv\psi(v)=K_0 \ .
\label{steadyintegral}
\EE
This is transformed into a differential-difference equation after
differentiation with respect to $u$:
\BE
a\psi'(u)+\psi(u+\sigma)-\psi(u-\sigma)=0\ , {\rm where}\
\psi(u)\neq 0 \ .
\label{diffdiff}
\EE
This steady equation has many solutions, including the {\sl
natural} one $\psi_0=K_0/(a+2\sigma)$ which is non-vanishing
everywhere, or delta combs such as (\ref{deltas}). We search for
solutions of the type in Fig. \ref{fig:analyticlumps}, i.e.
periodic arrays of lumps, of period $d$, each one having a
symmetric {\sl hump} shape $f(u)$ of width $2L$ (i.e. $f(u)=0$ if
$u \notin [-L,L]$):
\BE
\psi(u)=\sum_{n=-\infty}^\infty f(x-nd)
\label{sumoflumps}
\EE
We are assuming that the lumps do not overlap, so that $d>2L$. We
also note that if $\sigma+2L<d$ there is no interaction between
different lumps, so that for $u\in [-L,L]$ Eq. (\ref{diffdiff})
reduces to $f'(u)=0$ and there is no lump solution. Moreover,
analysis is much simplified if each of the lumps interacts only
with its neighbors ($\sigma+2L<2d$). Thus we restrict to
$d<\sigma+2L<2d$, for which (\ref{diffdiff}) with
(\ref{sumoflumps}) and $u \in (-L,L)$ becomes:
\BE
af'(u)+f(u+\sigma-d)-f(u-\sigma+d)=0
\EE
The general solution of this linear equation is obtained as a sum
of exponentials $\exp(\lambda u)$, with
\BE
a\lambda=\sinh\left(\lambda(d-\sigma)\right)\ .
\label{trascen}
\EE
$\lambda=0$ is always a solution, and if $d-\sigma<a$ there are
two additional solutions $\pm\lambda$, plotted in Fig.
\ref{fig:lambda}. For $d-\sigma>a$ the only solution is the
constant one, but in the opposite case (the situation favored by
enhanced intraspecific competition $a$) the solution is a linear
combination of three exponentials. Two of the constants of the
combination are determined from $f(L)=f(-L)=0$. The third one,
which gives the overall normalization, can be obtained by
returning back to the original equation (\ref{steadyintegral}).
Finally we get
\BA
f(u) &=& A \left( 1 -\frac{\cosh(\lambda u)}{\cosh(\lambda L)}
\right)  \  {\rm if}  \ u \in [-L,L] \nonumber \\
  &=& 0  \ \qquad {\rm elsewhere}
\label{fhump}
\EA
with
\BE
A=\frac{K_0}{a\left(1-{\rm sech}(\lambda
L)\right)+\frac{2}{\lambda}\left(\lambda L - \tanh(\lambda L)
\right)} \ ,
\label{normalization}
\EE
and the value of $\lambda$ which is plotted in Fig.
\ref{fig:lambda}. Figure \ref{fig:analyticlumps} shows the
analytic solution (\ref{sumoflumps}) with
(\ref{fhump})-(\ref{normalization}). We have not studied the
stability of this configuration. But the numerical results in
\citet{Pigolotti2007} indicate that it is stable for some values
of $L$ and $d$.

\begin{figure}[ht]
    \begin{center}
    \includegraphics[width=\columnwidth,clip]{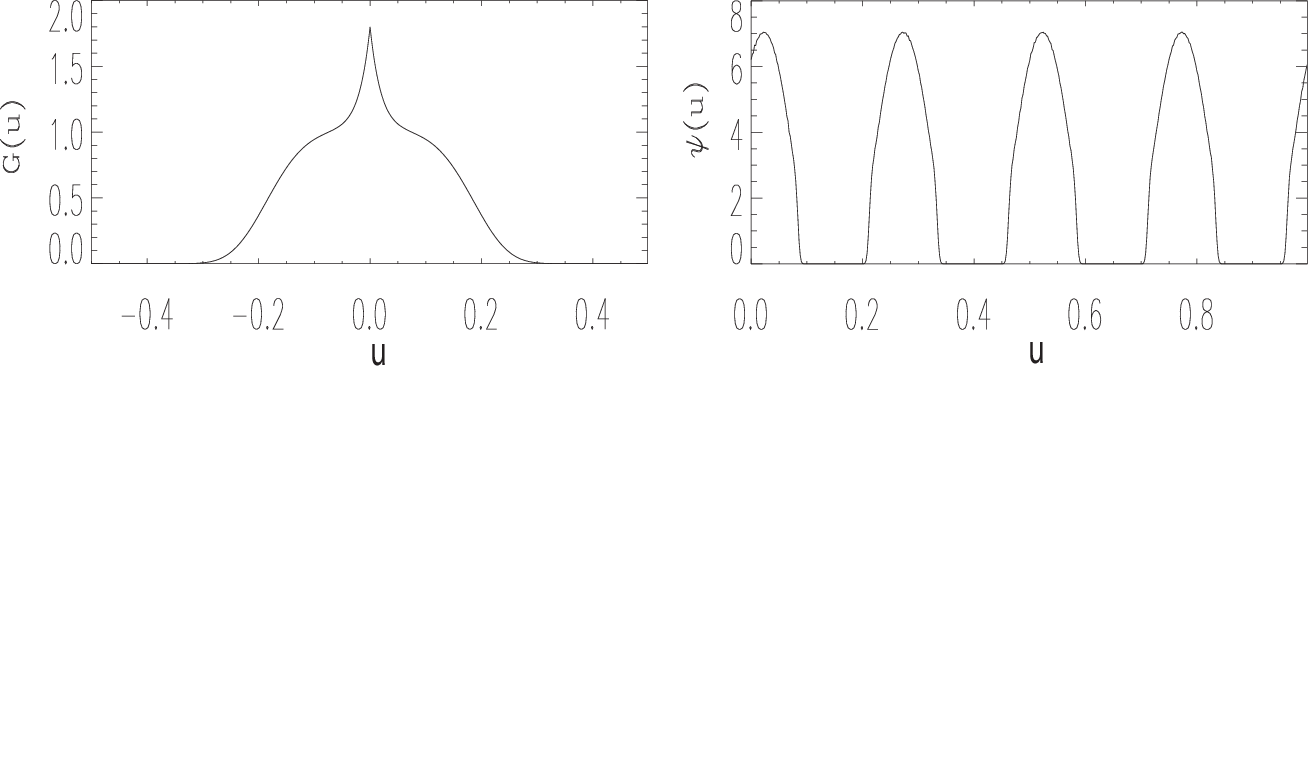}
    \end{center}
\caption{\label{fig:lumps} The kernel $G=g_{0.2}^4+0.8 g_{0.02}^1$
(left), and the steady solution obtained numerically from it at
long times with constant $K_0=1$ (right). }
\end{figure}

We finally stress that the appearance of the lumped solution is
not a consequence of the singularity of the delta function in the
kernel. In fact, any kernel sufficiently peaked at the origin will
favor the coexistence of close species. If the behavior at larger
distances of the kernel makes it not positive-definite, then full
coexistence will be unstable and the natural solution will split
into disjoint lumps. An example of the final steady state in this
situation is shown in Fig. \ref{fig:lumps}, with a kernel
$G=g_{0.2}^4+0.8 g_{0.02}^1$ which has the properties just
described and contains no delta singularity.

\begin{acknowledgements}
C.L. and E.H-G. acknowledge support from project FISICOS
(FIS2007-60327) of MEC and FEDER and NEST-Complexity project
PATRES (043268). K.H.A was supported by the Danish Research
Council, grant no. 272-07-0485
\end{acknowledgements}

\appendix{Models leading to LV competitive interactions}
\label{sec:models}

We have seen that the character of the interaction kernel $G$ is
of major importance to determine the qualitative outcome of LV
competition. In the original formulation of the niche model,
however, only positive definite kernels were allowed.  The reason
is that competition kernels were derived in terms of utilization
functions $u_i(x)$, describing how consumer $i$ uses resource at
niche location $x$ (assumed to be continuous)
\citep{MacArthur1967,Roughgarden1979}:
\begin{equation}
  \label{eq:interaction}
  G_{ij} = \frac{ \int u_i(x) u_j(x)\, dx }{ \int u_i^2(x)\, dx }.
\end{equation}
When the resource is directly related to space,
(\ref{eq:interaction}) can be justified by considering the
probability that consumer $i$ meets consumer $j$
\citep{Roughgarden1979}. It is easy to see that $G_{ij}$ obtained
from (\ref{eq:interaction}) is positive definite. We show in the
following, however, that relation (\ref{eq:interaction}) is by no
means general, and that a greater variety of kernels --positive or
non-positive definite, so that the natural solution representing
coexistence can be either stable or unstable-- could be obtained
from equations in which resources are explicitly modelled. Related
calculations could be found, for example, in \citet{Schoener1974}.

We consider a set of predators (or consumers), with populations
$N_i$, $i=1,2,...m$, competing for different types of prey
populations or resources, $R_\alpha$, $\alpha=1,2,...n$, the later
growing in a logistic way with growth rate $\beta_\alpha$ and
carrying capacity $Q_\alpha$ in the absence of predators.
Particular equations modelling this are
\BA
\dot R_\alpha &=& -R_\alpha \sum_i a_{\alpha i} N_i +\beta_\alpha
R_\alpha
\left(1-\frac{R_\alpha}{Q_\alpha}\right) \\
\dot N_i &=& N_i \sum_\alpha S_{i\alpha} R_\alpha -d_i N_i
\label{logisticfood}
\EA
$d_i$ is the death rate of species $i$. The interaction
coefficients are $a_{\alpha i}$, the depletion rate of resource
$\alpha$ produced by species $i$, and the {\sl sensitivity}
$S_{i\alpha}$, giving the growth rate of $i$ thanks to resource
$\alpha$. Lotka-Volterra type dynamics arises when the time scale
for resource evolution is much faster than that of the consumers
(i.e. $S_{i\alpha}$ and $d_i \rightarrow \infty$, but with their
ratio finite). In this case, adiabatic elimination of the resource
can be done ($\dot R_\alpha \approx 0$, so that each prey is at
each instant at the equilibrium determined by their consumers),
giving
\BE
R_\alpha \approx Q_\alpha \left( 1-\frac{1}{\beta_\alpha}\sum_i
a_{\alpha i} N_i\right)  \ .
\EE
for the non-vanishing resources. The {\sl impact} matrix,
$D_{\alpha i}$, describing the depletion of resource $\alpha$ by
species $i$ \citep{Meszena2006}, is $ D_{\alpha i}=Q_\alpha
a_{\alpha i}/\beta_\alpha$, which substituted into the consumers
equation leads to:
\BE
\dot N_i = N_i \left(r_i  - \sum_j C_{ij} N_j \right) \ ,
\label{effectiveLV1}
\EE
where $r_i=\sum_\alpha S_{i\alpha} Q_\alpha$ is the maximum growth
rate and $C_{ij}=\sum_\alpha S_{i\alpha} D_{\alpha j}$.
Thus, the result is an effective interaction among the predators which
is of Lotka-Volterra type. It is customary to write
(\ref{effectiveLV1}) in terms of the carrying capacity $K_i$,
defined as the equilibrium population $N_i$ attained in the
absence of the other competitors, i.e. $K_i = r_i/C_{ii}$. In
terms of it, Eq. (\ref{effectiveLV1}) becomes identical to
(\ref{LV}), with
\BE
G_{ij}=\frac{C_{ij}}{C_{ii}} = \frac{\sum_\alpha S_{i\alpha}
D_{\alpha j}}{\sum_\alpha S_{i\alpha} D_{\alpha i}}.
\label{Gij}
\EE

Having a continuum $R(x)$ of resources instead of a discrete set
$R_\alpha$  does not introduce important difficulties. Simply one
should replace sums by integrals, replacing the coefficients of Eq. (\ref{effectiveLV1}) by:
\BA
r_i &=& \int   S_i(x) Q(x) dx \ , \label{effectiveCont1} \\
C_{ij} &=& \int  S_i(x) D_j(x) d x \ , \label{effectiveCont2} \\
G_{ij} &=& \frac{\int  S_i(x) D_j(x) dx}{\int S_i(x) D_i(x) dx } \
,
\label{effectiveCont3}
\EA
One can also consider a continuum of species, labelled by their
phenotypes $u$, so that Eq. (\ref{LV}) is replaced by Eq.
(\ref{LVcon}) with $K(u) = r(u)/C(u,u)$, $G(u,v)=C(u,v)/C(u,u)$,
and $r(u)$ and $C(u,v)$ given by obvious generalizations of
(\ref{effectiveCont1}) and (\ref{effectiveCont2}).

It is clear that the presence in the kernel $G_{ij}$ of two
different functions (compare with the most restrictive expression
(\ref{eq:interaction})) gives enough freedom to obtain a variety
of kernel behaviors under different circumstances. A particularly
popular choice is to assume that impact and sensitivity are
proportional: $S_{i\alpha}=\epsilon D_{\alpha i}$, with a constant
efficiency $\epsilon$. In the continuum resource case the
functions can be written in terms of a single utilization function
$u_i(x)$ as $D_i(x)=u_i(x)$ and $S_i(x)=\epsilon u_i(x)$, leading
to the classical expression (\ref{eq:interaction}). Slightly more
general cases arise when the efficiency depends only on the
resource, $\epsilon=\epsilon_\alpha$, or on the consumer
$\epsilon=\epsilon_i$, or when dependence on the two types of
species factorizes, $\epsilon=v_i w_\alpha$. In all these cases
(if the efficiency is positive) one is lead to a kernel $G_{ij}$
which is positive definite. In more general cases, one can
have a kernel leading to instability of the coexistence state.

We conclude with two
instances of ecological interactions in (\ref{logisticfood})
which allow to tune the stability.
First, a homogeneous discrete and infinite niche space in which
all resources have the same internal dynamics $Q_\alpha=Q$,
$\beta_\alpha=\beta$, $\forall \alpha$, as well as the consumers:
$d_i=d$, $\forall i$. The interactions are taken to be
\BA
S_{i\alpha}&=& g \delta_{i,\alpha} \\
a_{\alpha, j}= \frac{\beta}{Q} D_{\alpha j} &=& a\delta_{\alpha,j}
+b (\delta_{\alpha, j-1}+\delta_{\alpha, j+1}).\nonumber
\EA
This models a situation in which the consumer $k$ grows only by
consuming its optimal resource $R_k$, whereas it depletes also the
neighboring resources, $R_{k+1}$ and $R_{k-1}$. We have $r_i=Qg$,
$K_i=\beta/a$, $C_{ij}=(Qg/\beta)\left(a\delta_{i,j} +b
(\delta_{i, j-1}+\delta_{i, j+1})\right)$, and
$G_{ij}=\delta_{ij}+(b/a)\left(\delta_{i, j-1}+\delta_{i,
j+1}\right)$ so that equation (\ref{LV}) is now
\BE
\dot N_i = Q g N_i \left[ 1 -\frac{1}{\beta} \left(  a N_i +b
\left( N_{i+1}+N_{i-1}\right)\right) \right].
\EE
The natural solution, i.e. the one in which all species have
positive non zero population, is $\overline N_i=\beta/(a+2b)$,
$\forall i$. Its linear stability can be studied by linearization,
$N_l(t)=\overline N_l + \delta N_l(t)$ and substitution of the
ansatz $\delta N_l \approx e^{\lambda_q t} e^{i q l}$ (here
$i=\sqrt{-1}$). We find $\lambda_q=-(Qg/\beta)\left( a+2b\cos q
\right)$, $q\in [-\pi,\pi]$. $\lambda_q$ are the eigenvalues of
$-C_{ij}$, and stability of $\overline N_i$ requires all these
eigenvalues to be negative, i.e., $C_{ij}$ to be positive
definite. When $a> 2b$, then $\lambda_q<0$ $\forall q$, and the
natural coexistence solution is globally stable (see results in
Sect. \ref{sec:stability}). It is unstable otherwise. In this
example, there is no well-defined single utilization function and
the positivity properties of the interaction kernel and thus the
stability of the natural solution can be changed by varying the
parameters.

As a second example, with nonconstant carrying capacity, we
consider a continuous distribution of resources and species on the
line, and we take
\BE
\begin{array}{rclrcl}
Q(x)  &=& Q g(x),     & \beta(x) &=& \beta g(x), \nonumber \\
S_u(x)&=& s\delta(u-x),    & a_u(x)   &=& f(u-x).
\end{array}
\EE
which implies that the consumers of phenotype $u$ grow only from
the resource at location $x=u$, but they deplete a wider range
characterized by $f$. This leads (in the continuous formalism) to
$r(u) = s Q g(u)$ , $C(u,v)= s Q f(u-v)/\beta$, $K(u) = \beta
g(u)/f(0)$ and $G(u,v)=f(u-v)/f(0)$. In this example, by choosing
the functions $g$ and $f$,  we can impose any desired combinations
of carrying capacity and interaction kernel. Gaussianity or
positive definiteness are particular cases, no more natural in
this generalization than alternative choices leading to
non-positiveness, instability, and thus exclusion zones between
clumps of species.


\label{lastpage}

\end{document}